\documentclass[12pt]{article}
\RequirePackage[colorlinks,citecolor=blue,urlcolor=blue]{hyperref}
\input{hzz_rjournal.sty}
\begin{document}
\title{hdtg: An R package for high-dimensional truncated normal simulation}
\author{Zhenyu Zhang$^{1}$,\\
	 Andrew Chin$^{2}$, \\
	 Akihiko Nishimura$^{2}$, \\
	  and Marc A.~Suchard$^{1,3,4}$}
\maketitle
\noindent $^1$\textit{Department of Biostatistics,  University of California Los Angeles} \\
$ ^2 $\textit{Department of Biostatistics, Johns Hopkins University }\\
$^3$\textit{Department of Biomathematics}, $^4$\textit{Department of Human Genetics, Universtiy of \\ California Los Angeles} \\

\begin{abstract}
Simulating from the multivariate truncated normal distribution (MTN) is required in various statistical applications yet remains challenging in high dimensions.
Currently available algorithms and their implementations often fail when the number of parameters exceeds a few hundred.
To provide a general computational tool to efficiently sample from high-dimensional MTNs, we introduce the \href{https://cran.r-project.org/package=hdtg}{hdtg} package that implements two state-of-the-art simulation algorithms: harmonic Hamiltonian Monte Carlo (\hhmc) and zigzag Hamiltonian Monte Carlo (\zhmc). 
Both algorithms exploit analytical solutions of the Hamiltonian dynamics under a quadratic potential energy with hard boundary constraints, leading to rejection-free methods.
We compare their efficiencies against another state-of-the-art algorithm for MTN simulation, the minimax tilting accept-reject sampler (\mt).
The run-time of these three approaches heavily depends on the underlying multivariate normal correlation structure.
\zhmc and \hhmc both achieve 100 effective samples within 3,600 seconds across all tests with dimension ranging from 100 to 1,600, while \mt has difficulty in several high-dimensional examples.
We provide guidance on how to choose an appropriate method for a given situation and illustrate the usage of \href{https://cran.r-project.org/package=hdtg}{hdtg}.
\end{abstract}
	%\doublespacing \todo{no double spacing for submission}

\section{Introduction}
Sampling from a multivariate truncated normal (MTN) distribution is a recurring problem in many statistical applications. 
The MTN distribution of a $ \dimd $-dimensional random vector $ \bm{x} \in \realNumbers^{\dimd}$ has the form 
\begin{equation}\label{eq:mtn}
\bm{x} \sim \normalDensityFunction{\mean}{\covariance} \text{ with } \lb \leq \bm{x} \leq \ub \text{ bounded}, 
\end{equation}
where $ \mean $ and $ \covariance $ are the mean vector and covariance matrix, and $ \lb, \ub \in \realNumbers^{\dimd}$ denote the lower and upper truncation bounds.
MTNs arise in various context including probit and tobit models \citep{albert1993bayesian, tobin1958estimation}, latent Gaussian models \citep{bolin2015excursion}, copula regression \citep{pitt2006efficient}, spatial models \citep{tsionas2016spatial, baltagi2018generalized, zareifard2021heterogeneous}, Bayesian metabolic flux analysis \citep{heinonen2019bayesian}, and many others.
When the dimension $ \dimd $ is small, a standard rejection sampler \citep{geweke1991efficient, kotecha1999gibbs} works well and is a common choice.
However, simulation from a larger MTN with hundreds or thousands of correlated dimensions remains a computational challenge.
Work towards this goal include harmonic Hamiltonian Monte Carlo \citep[\hhmc]{pakman2014exact}, rejection sampling based on minimax (saddle point) exponential tilting \citep[\mt]{botev2017normal}, and the most recent Zigzag Hamiltonian Monte Carlo \citep[\zhmc]{nishimura2020discontinuous, nishimura2021hamiltonian} methods.

The \mt method provides independent samples but can suffer from low acceptance rates and becomes impractical with $ \dimd > 100 $, except in special cases like when the MTN has a strongly positive correlation structure \citep{botev2017normal}.
Both \hhmc and \zhmc are Markov chain Monte Carlo (MCMC) approaches that generate correlated samples, but can nonetheless be highly efficient and scale to thousands or more dimensions.  
To our knowledge, however, there is no general-purpose implementation of either method;
the \texttt{tmg} package provided by \citet{pakman2014exact} is no longer available on CRAN, and \citet{zhang2022hamiltonian} implement \zhmc for their phylogenetics applications in the specialized software BEAST \citep{beast2018}. 
Therefore, we have developed the \href{https://cran.r-project.org/package=hdtg}{hdtg} R package for efficient  MTN simulation.
The package implements tuning-free \zhmc and \hhmc.
We provide performance comparisons among these two methods and a \mt implementation from the \href{https://cran.r-project.org/package=TruncatedNormal}{TruncatedNormal} package \citep{tnpkg}.
In most of the test cases with $ \dimd > 100 $, \hhmc and \zhmc outperform MET. 
We then conclude with some empirical guidance on which method to use in different scenarios.

\section{Algorithm}
\label{sec:algorithm}
We begin by briefly introducing \hhmc and \zhmc, both of which are variants of HMC, an effective proposal generation mechanism exploiting the properties of Hamiltonian dynamics \citep{hmcneal}.
\hhmc and \zhmc follow the same general framework.
To sample $ \bposition = \left(\position_1, \dots, \position_\dimd\right) \in \realNumbers^{\dimd}$ from the target distribution $ \pi(\bposition) $, the HMC variants introduce an auxiliary \textit{momentum} variable $ \bmomentum$ and define an augmented target distribution $\pi(\bposition, \bmomentum) = \pi(\bposition) \pi(\bmomentum)$ in the joint space.
They then propose the next state by first re-sampling the momentum variable from its marginal and then simulating the solution of Hamiltonian dynamics governed by the differential equations
\begin{equation}
\label{eq:hamilton}
\frac{\diff \bposition}{\diff t}
= \nabla \kinetic(\bmomentum), \quad
\frac{\diff \bmomentum}{\diff t}
= - \nabla \potential(\bposition),
\end{equation}
where $\potential(\bposition)=- \log \pi(\bposition)$ and $\kinetic(\bmomentum) = - \log \pi(\bmomentum)$ are referred to as \textit{potential} and \textit{kinetic} energies.
The dynamics are simulated for a pre-set time duration $\integrationTime$ and the end state constitutes a valid Metropolis proposal to be accepted or rejected according to the standard formula \citep{metropolis53, hastings1970monte}.

The most common versions of HMC use the momentum distribution $ \momentumDistribution{\bmomentum} \sim \mathcal{N}(\bm{0}, \M) $ and rely on the leapfrog integrator to numerically solve \eqref{eq:hamilton}, as its solutions are analytically intractable in general settings. \hhmc takes advantage of the fact that \eqref{eq:hamilton} admits analytical solutions when the target $\pi(\bposition)$ is an MTN.
The solution follows independent harmonic oscillations along the principal components of the covariance/precision matrix \citep{pakman2014exact};
we thus refer to the algorithm as \hhmc. 
Truncation boundaries are handled via elastic ``bounces'' against hard ``potential energy walls''  \citep{hmcneal}.
We refer interested readers to \citet{pakman2014exact} for details on \hhmc.

\zhmc differs from the common HMC versions in that it deploys a Laplace momentum  \citep{nishimura2020discontinuous, nishimura2021hamiltonian} 
\begin{equation}
\momentumDistribution{\bmomentum} \propto \prod_{i} \exp\left(-|\momentum_i|\right), i = 1, \dots, \dimd.
\end{equation}
The Hamiltonian dynamics then become
\begin{equation}
\label{eq:hzz_equation}
\frac{{\rm d} \bposition}{{\rm d} t}
=   \sign\left(\bmomentum\right), \quad
\frac{{\rm d} \bmomentum}{{\rm d} t} = - \nabla \potential(\bposition),
\end{equation}
where $\sign\left(\momentum_i\right)$ returns 1 if $\momentum_i$ is positive and -1 otherwise.
Because the velocity $\diff \bposition / \diff t \in \{\pm 1\}^\dimd$ remains constant until one of the $ \momentum_i $ flips its sign, the trajectory of these Hamiltonian dynamics has a zigzag pattern, hence the name \zhmc.
The zigzag dynamics also admit analytical solutions under an MTN target and can handle the truncation in the same manner as in \hhmc.
We refer interested readers to \citet{nishimura2021hamiltonian} and \citet{zhang2022hamiltonian} for \zhmc algorithm details, including how to determine the time of a momentum sign change and of a bounce against truncation boundaries. 

The simulation duration $\integrationTime$, i.e.\ how long Hamiltonian dynamics is simulated for each proposal generation, critically affects efficiencies of both Harmonic and Zigzag-HMC.
For \hhmc, \citet{pakman2014exact} suggest setting $ \integrationTime = \pi/2 $;
when using this fixed $ \integrationTime $, however, we observe inefficiencies in some of our examples in Section~\ref{sec:efficiency_comparison} due to Hamiltonian dynamics' periodic behaviors \citep{hmcneal}.
We therefore randomize the duration $\integrationTime$, as recommended by \citet{hmcneal}, and draw it from a uniform distribution on $ \left[\pi/8,\pi/2\right] $. 
For \zhmc, we adopt the choice $ \integrationTime =  \sqrt{2} \eigenvalue{\text{min}}^{-1/2}$ based on the heuristics of \cite{nishimura2021hamiltonian}, where $\eigenvalue{\text{min}} $ is the minimal eigenvalue of the precision matrix $ \precision = \covariance^{-1} $.
We compute $\eigenvalue{\text{min}} $ using the Lanczos algorithm \citep{demmel1997applied} as in the \href{https://cran.r-project.org/package=mgcv}{mgcv} package \citep{mgcvcite}.
We further implement the no-U-turn algorithm (NUTS) of \citet{hoffman2014nuts} to automatically determine
the integration time. 
With NUTS, we only need to pick a base integration time $ \basestep $ which we set to $0.1 \eigenvalue{\text{min}}^{-1/2}$
as recommended by \citet{nishimura2021hamiltonian}.

\section{Using hdtg}\label{sec:pkg}
The \href{https://cran.r-project.org/package=hdtg}{hdtg} package allows users to draw MCMC samples from an MTN with fixed or random mean and covariance/precision matrix.
In our current implementation, \zhmc accepts the most commonly seen element-wise truncations as in Equation \eqref{eq:mtn} while \hhmc can handle a more general constraint  
\begin{equation}
	\left(\fmat\bposition + \gvec\right)_i \geq 0, \text{ for } i = 1, \dots, \numTruncs.
\end{equation}
Here the $ \numTruncs \times \dimd $ matrix $ \fmat $ and $ \numTruncs$-dimensional vector $ \gvec $ specify the truncations and $ \left(\cdot\right)_i $ denotes the $ i$th vector element. 
As an example, one may use the following code to generate 1,000 samples from a 10-dimensional MTN with zero mean and an identity covariance matrix truncated to the positive orthant:

\begin{example}
# set the random seed
set.seed(1)
# draw MTN samples using Zigzag-HMC 
samplesZHMC <- zigzagHMC(n = 1000, mean = rep(0, 10), prec = diag(10), 
                         init = rep(0.1, 10), lowerBounds = rep(0, 10), 
                         upperBounds = rep(Inf, 10))
# draw MTN samples using Harmonic-HMC 
samplesHHMC <- harmonicHMC(n = 1000, mean = rep(0, 10), 
		           choleskyFactor = diag(10), precFlg = TRUE, 
		           init = rep(0.1, 10),  F = diag(10), g = rep(0, 10))
\end{example}
The arguments are:
\begin{itemize}
	\item \code{n}: number of samples.
	\item \code{mean}: a $ \dimd $-dimensional mean vector.
	\item \code{prec}: the precision matrix.
	\item \code{init}: a vector of the initial value that must satisfy all constraints.
	\item \code{lowerBounds}: a $ \dimd $-dimensional vector specifying the lower bounds.
	\item \code{upperBounds}: a $ \dimd $-dimensional vector specifying the upper bounds.
	\item \code{choleskyFactor}: upper triangular matrix $ \umat $ from Cholesky decomposition of precision or covariance matrix into $ \umat\transpose\umat $.
	\item \code{precFlg}: whether \code{choleskyFactor} is from precision (\code{TRUE}) or covariance matrix (\code{FALSE}).
	\item \code{F}: the $ \fmat $ matrix.
	\item \code{g}: the $ \gvec $ vector.
	
\end{itemize}
%Three things  function \textbf{rmtn} SIMD (single instruction-stream, multiple data-stream)
With a random $ \mean $ or $ \precision $, one can simply call \code{zigzagHMC} or \code{harmonicHMC} and pass the updated $ \mean $ and $ \precision $ as arguments.
But a more efficient usage of \zhmc exists. \code{zigzagHMC} calls the function \code{createEngine} (or \code{createNutsEngine} if using NUTS) to create a C\texttt{++} object that sets up truncation boundaries and SIMD (single instruction-stream, multiple data-stream) vectorization.
Therefore, we can avoid repeated calls of \code{createEngine} by reusing the C\texttt{++} object, as in the following example where the 10-dimensional target MTN has a random mean and precision:
\begin{example}
set.seed(1)
n <- 1000 
d <- 10
samples <- array(0, c(n, d))
# initialize MTN mean and precision
m <- rnorm(d, 0, 1)
prec <- rWishart(n = 1, df = d, Sigma = diag(d))[,,1]

# call createEngine once
engine <- createEngine(dimension = d, lowerBounds = rep(0, d),
upperBounds = rep(Inf, d), seed = 1, mean = m, precision = prec)

HZZtime <- sqrt(2) / sqrt(min(mgcv::slanczos(A = prec, k = 1, 
			  kl = 1)[['values']]))

currentSample <- rep(0.1, d)
for (i in 1:n) {
   m <- rnorm(d, 0, 1)
   prec <- rWishart(n = 1, df = d, Sigma = diag(d))[,,1]
   setMean(sexp = engine$engine, mean = m)
   setPrecision(sexp = engine$engine, precision = prec)
   currentSample <- getZigzagSample(position = currentSample, nutsFlg = F,
	   		         engine = engine, stepZZHMC = HZZtime)
   samples[i, ] <- currentSample
}
\end{example}

\section{Efficiency comparison and method choice}
\label{sec:efficiency_comparison}
To assess the performance of \hhmc, \zhmc and \mt, we compare them on MTNs with a variety of correlation structures. 
The three examples are: 1) MTNs with its covariance matrix $ \covariance $ drawn from the uniform LKJ distribution \citep{lewandowski2009generating} as implemented in the \code{rlkjcorr} function from package \href{https://cran.r-project.org/package=trialr}{trialr} \citep{trialrpkg};
2) MTNs with a compound symmetric covariance matrix such that $ \covariance_{i,i} = 1 $ and $ \covariance_{i,j} =  0.9 $ for $ i \neq j $; and
3) a real-world MTN that arises as a posterior conditional distribution in a statistical phylogenetics model of HIV evolution \citep{zhang2021large, zhang2022hamiltonian}.
For simplicity, we assume the truncation $ x_i > 0 $ for $ i = 1, \dots, \dimd$ in the first two examples.
For the HIV example, the truncation is determined by the signs of observed binary biological features.

We now specify our comparison criteria and the rationale behind them.
A more efficient MCMC algorithm takes shorter time to achieve a certain effective sample size (ESS).
For all three samplers considered, we compare their run-time to obtain the first one or 100 effectively independent samples ($\timeFirstSample$ and $ \timeHundredSamples $).
We include both $\timeFirstSample$ and $\timeHundredSamples$ because $\timeHundredSamples $ reflects a practical run-time for simulation from a fixed MTN and $ \timeFirstSample $ better captures the pre-processing overhead that remains relevant in cases where $\covariance$ is random. 
Recall that the main pre-processing costs are the Cholesky decomposition of $ \covariance $ or $ \precision $ (\hhmc), calculating the minimal precision matrix eigenvalue $ \eigenvalue{\text{min}} $ (\zhmc), and solving the minimax optimization problem (\mt).
Therefore we have 
\begin{equation}\label{eq:time}
\begin{split}
\timeFirstSample & = \timePre + \timeIter \\
\timeHundredSamples & =  \timePre + 100\timeIter,
\end{split}
\end{equation}
where $ \timePre $ and $ \timeIter $ are the pre-processing time required for each $ \covariance $ update and the average run-time per one effective sample.
For simulation from a fixed MTN, $ \timePre $ is a one-time cost and so $ \timeHundredSamples $ serves as a better efficiency criterion.
When $ \cov $ is random (e.g.\ the second example in Section \ref{sec:pkg}), if $ \cov $ changes its value $ k $ times, the total run-time to obtain one effective sample for each $ \cov $ is $ k \timeFirstSample $ and so the $ \timeFirstSample $ criterion would be more informative.

For \hhmc and \zhmc, we estimate the ESS using the \href{https://cran.r-project.org/package=coda}{coda} package \citep{codapkg} and define $ \nsample $ as the average number of MCMC iterations required for one effectively independent sample.
We approximate $ \nsample $ by $ \chainLength / \text{ESS}_{\text{min}}$, where $\text{ESS}_{\text{min}}$ is the minimal ESS across all dimensions and $\chainLength $ is the chain length. 
We fix $ \nsample = 1$ for \mt as it generates independent samples.
Therefore $ \timeIter $ in Equation \eqref{eq:time} equals the average time to complete $ \nsample $ iterations after pre-processing.
Table \ref{tb:All} reports our efficiency comparison in terms of $\timeFirstSample$ and $ \timeHundredSamples $.
We run each test on a quad-core Intel i7 4 GHZ equipped machine with 32GB of memory.

\tbAll

\par

The efficiency of all three methods strongly depends on the correlation structure.
\mt fails to generate 100 effectively independent samples within two hours in a few higher dimensional tests, while \hhmc and \zhn enjoy a $ \timeHundredSamples < 3600$ seconds across all tests.
In the LKJ example, \zhn become more efficient than \hhmc when $ \dimd$ reaches $1600$.
\zhmc and \znuts tend to share similar performance.
While \znuts is the most efficient choice for the LKJ test ($ \dimd = 1600 $), \zhmc wins the test on an MTN from the HIV example ($ \dimd = 800, 1600 $). 
On the other hand, when $ \covariance $ is compound symmetric with a high correlation of 0.9, \hhmc consistently outperforms the other methods.
When \mt does function for a target MTN, its $ \timeHundredSamples $ is close to $ \timeFirstSample $, as solving the initial minimax optimization problem takes most of its run-time.   

In practice, we recommend running a quick efficiency comparison to decide which method to use.
Nevertheless we provide some general guidance on method choice for high-dimensional MTN simulation:
\begin{itemize}
	\item If $ \dimd \leq 100 $ or the correlation structure is strongly positive, use \mt or \hhmc. 
	\hhmc may run faster but \mt has the advantage of generating independent samples.
	\item For all other cases,  \zhn is presumably more efficient, although \hhmc may outperform them when $ \dimd < 1000$. 
	\item It is always worth trying \mt which is free of MCMC convergence concerns. Since our simulation only examines a few correlation structures, it is possible that \mt can handle other large MTNs. 
\end{itemize}
A final point that needs consideration is that \zhn require $ \precision $ and if only $ \covariance $ is available, the method first inverts $ \covariance $. This is a one-time operation and likely negligible cost when $ \covariance $ is constant. 
The approaches does become expensive if $ \covariance $ is random, as the $ \order{\dimd^3} $ inversion is necessary for each value of $ \covariance $.
In practice, statistical models may be parameterized in terms of  $ \covariance $  \citep{lachaab2006modeling, molstad2021gaussian} or $ \precision $ \citep{baltagi2018generalized, lehnert2019large, li2020using}.
\hhmc carries a similar limitation since it requires a $ \order{\dimd^3} $ Cholesky decomposition of $ \covariance $ or $ \precision $, whichever is provided.
Therefore, when $ \dimd $ is large and the target MTN has a random correlation structure, one may favor \zhn over \hhmc especially if a closed-form $ \precision $ is at hand.

\section{Conclusion}
This article introduces the \href{https://cran.r-project.org/package=hdtg}{hdtg} package oriented for efficient MTN simulation.
In most of our high-dimensional tests the implemented \hhmc and \zhmc algorithms outperform the current best approach available in the \href{https://cran.r-project.org/package=TruncatedNormal}{TruncatedNormal} package.
To our best knowledge, \href{https://cran.r-project.org/package=hdtg}{hdtg} is the first tool that can generate samples from an arbitrary MTN with thousands of dimensions. 
We discuss the usage of functions and provide practical suggestions on method choice.
We expect to see future large-scale statistical applications utilizing the efficiency of \href{https://cran.r-project.org/package=hdtg}{hdtg}.

\section{Funding}

This work was partially supported through National Institutes of Health grant R01 AI153044.

\bibliographystyle{chicago}
\bibliography{hzz_rjournal}

\begin{thebibliography}{}

\bibitem[\protect\citeauthoryear{Albert and Chib}{Albert and
  Chib}{1993}]{albert1993bayesian}
Albert, J.~H. and S.~Chib (1993).
\newblock Bayesian analysis of binary and polychotomous response data.
\newblock {\em Journal of the American statistical Association\/}~{\em
  88\/}(422), 669--679.

\bibitem[\protect\citeauthoryear{Baltagi, Egger, and Kesina}{Baltagi
  et~al.}{2018}]{baltagi2018generalized}
Baltagi, B.~H., P.~H. Egger, and M.~Kesina (2018).
\newblock Generalized spatial autocorrelation in a panel-probit model with an
  application to exporting in {C}hina.
\newblock {\em Empirical Economics\/}~{\em 55\/}(1), 193--211.

\bibitem[\protect\citeauthoryear{Bolin and Lindgren}{Bolin and
  Lindgren}{2015}]{bolin2015excursion}
Bolin, D. and F.~Lindgren (2015).
\newblock Excursion and contour uncertainty regions for latent {G}aussian
  models.
\newblock {\em Journal of the Royal Statistical Society: Series B (Statistical
  Methodology)\/}~{\em 77\/}(1), 85--106.

\bibitem[\protect\citeauthoryear{Botev and Belzile}{Botev and
  Belzile}{2021}]{tnpkg}
Botev, Z. and L.~Belzile (2021).
\newblock {\em TruncatedNormal: Truncated Multivariate Normal and Student
  Distributions}.
\newblock R package version 2.2.2.

\bibitem[\protect\citeauthoryear{Botev}{Botev}{2017}]{botev2017normal}
Botev, Z.~I. (2017).
\newblock The normal law under linear restrictions: simulation and estimation
  via minimax tilting.
\newblock {\em Journal of the Royal Statistical Society: Series B (Statistical
  Methodology)\/}~{\em 79\/}(1), 125--148.

\bibitem[\protect\citeauthoryear{Brock}{Brock}{2020}]{trialrpkg}
Brock, K. (2020).
\newblock {\em trialr: Clinical Trial Designs in 'rstan'}.
\newblock R package version 0.1.5.

\bibitem[\protect\citeauthoryear{Demmel}{Demmel}{1997}]{demmel1997applied}
Demmel, J.~W. (1997).
\newblock {\em Applied numerical linear algebra}.
\newblock SIAM.

\bibitem[\protect\citeauthoryear{Geweke}{Geweke}{1991}]{geweke1991efficient}
Geweke, J. (1991).
\newblock Efficient simulation from the multivariate normal and student-t
  distributions subject to linear constraints and the evaluation of constraint
  probabilities.
\newblock In {\em Computing science and statistics: Proceedings of the 23rd
  symposium on the interface}, pp.\  571--578. Citeseer.

\bibitem[\protect\citeauthoryear{Hastings}{Hastings}{1970}]{hastings1970monte}
Hastings, W.~K. (1970).
\newblock {Monte Carlo} sampling methods using {Markov} chains and their
  applications.
\newblock {\em Biometrika\/}.

\bibitem[\protect\citeauthoryear{Heinonen, Osmala, Mannerstr{\"o}m, Wallenius,
  Kaski, Rousu, and L{\"a}hdesm{\"a}ki}{Heinonen
  et~al.}{2019}]{heinonen2019bayesian}
Heinonen, M., M.~Osmala, H.~Mannerstr{\"o}m, J.~Wallenius, S.~Kaski, J.~Rousu,
  and H.~L{\"a}hdesm{\"a}ki (2019).
\newblock Bayesian metabolic flux analysis reveals intracellular flux
  couplings.
\newblock {\em Bioinformatics\/}~{\em 35\/}(14), i548--i557.

\bibitem[\protect\citeauthoryear{Hoffman and Gelman}{Hoffman and
  Gelman}{2014}]{hoffman2014nuts}
Hoffman, M.~D. and A.~Gelman (2014).
\newblock The {No-U-Turn} sampler: adaptively setting path lengths in
  {H}amiltonian {M}onte {C}arlo.
\newblock {\em Journal of Machine Learning Research\/}~{\em 15\/}(1),
  1593--1623.

\bibitem[\protect\citeauthoryear{Kotecha and Djuric}{Kotecha and
  Djuric}{1999}]{kotecha1999gibbs}
Kotecha, J.~H. and P.~M. Djuric (1999).
\newblock Gibbs sampling approach for generation of truncated multivariate
  {G}aussian random variables.
\newblock In {\em 1999 IEEE International Conference on Acoustics, Speech, and
  Signal Processing. Proceedings. ICASSP99 (Cat. No. 99CH36258)}, Volume~3,
  pp.\  1757--1760. IEEE.

\bibitem[\protect\citeauthoryear{Lachaab, Ansari, Jedidi, and Trabelsi}{Lachaab
  et~al.}{2006}]{lachaab2006modeling}
Lachaab, M., A.~Ansari, K.~Jedidi, and A.~Trabelsi (2006).
\newblock Modeling preference evolution in discrete choice models: {A}
  {B}ayesian state-space approach.
\newblock {\em Quantitative Marketing and Economics\/}~{\em 4\/}(1), 57--81.

\bibitem[\protect\citeauthoryear{Lehnert, Kolbitsch, W{\"u}bbeler, Chiribiri,
  Schaeffter, and Elster}{Lehnert et~al.}{2019}]{lehnert2019large}
Lehnert, J., C.~Kolbitsch, G.~W{\"u}bbeler, A.~Chiribiri, T.~Schaeffter, and
  C.~Elster (2019).
\newblock {Large-scale Bayesian spatial-temporal regression with application to
  cardiac MR-perfusion imaging}.
\newblock {\em SIAM Journal on Imaging Sciences\/}~{\em 12\/}(4), 2035--2062.

\bibitem[\protect\citeauthoryear{Lewandowski, Kurowicka, and Joe}{Lewandowski
  et~al.}{2009}]{lewandowski2009generating}
Lewandowski, D., D.~Kurowicka, and H.~Joe (2009).
\newblock Generating random correlation matrices based on vines and extended
  onion method.
\newblock {\em Journal of multivariate analysis\/}~{\em 100\/}(9), 1989--2001.

\bibitem[\protect\citeauthoryear{Li, McComick, and Clark}{Li
  et~al.}{2020}]{li2020using}
Li, Z.~R., T.~H. McComick, and S.~J. Clark (2020).
\newblock {Using Bayesian latent Gaussian graphical models to infer symptom
  associations in verbal autopsies}.
\newblock {\em Bayesian analysis\/}~{\em 15\/}(3), 781.

\bibitem[\protect\citeauthoryear{Metropolis, Rosenbluth, Rosenbluth, Teller,
  and Teller}{Metropolis et~al.}{1953}]{metropolis53}
Metropolis, N., A.~W. Rosenbluth, M.~N. Rosenbluth, A.~H. Teller, and E.~Teller
  (1953).
\newblock Equation of state calculations by fast computing machines.
\newblock {\em Journal of Chemical Physics\/}~{\em 21\/}(6), 1087--1092.

\bibitem[\protect\citeauthoryear{Molstad, Hsu, and Sun}{Molstad
  et~al.}{2021}]{molstad2021gaussian}
Molstad, A.~J., L.~Hsu, and W.~Sun (2021).
\newblock Gaussian process regression for survival time prediction with
  genome-wide gene expression.
\newblock {\em Biostatistics\/}~{\em 22\/}(1), 164--180.

\bibitem[\protect\citeauthoryear{Neal}{Neal}{2011}]{hmcneal}
Neal, R.~M. (2011).
\newblock {MCMC} using {Hamiltonian} dynamics.
\newblock In S.~Brooks, A.~Gelman, G.~L. Jones, and X.-L. Meng (Eds.), {\em
  Handbook of Markov Chain Monte Carlo}, Volume~2. CRC Press New York, NY.

\bibitem[\protect\citeauthoryear{Nishimura, Dunson, and Lu}{Nishimura
  et~al.}{2020}]{nishimura2020discontinuous}
Nishimura, A., D.~B. Dunson, and J.~Lu (2020).
\newblock Discontinuous {H}amiltonian {M}onte {C}arlo for discrete parameters
  and discontinuous likelihoods.
\newblock {\em Biometrika\/}~{\em 107\/}(2), 365--380.

\bibitem[\protect\citeauthoryear{Nishimura, Zhang, and Suchard}{Nishimura
  et~al.}{2021}]{nishimura2021hamiltonian}
Nishimura, A., Z.~Zhang, and M.~A. Suchard (2021).
\newblock Hamiltonian zigzag sampler got more momentum than its markovian
  counterpart: Equivalence of two zigzags under a momentum refreshment limit.
\newblock {\em arXiv preprint arXiv:2104.07694\/}.

\bibitem[\protect\citeauthoryear{Pakman and Paninski}{Pakman and
  Paninski}{2014}]{pakman2014exact}
Pakman, A. and L.~Paninski (2014).
\newblock Exact {H}amiltonian {M}onte {C}arlo for truncated multivariate
  {G}aussians.
\newblock {\em Journal of Computational and Graphical Statistics\/}~{\em
  23\/}(2), 518--542.

\bibitem[\protect\citeauthoryear{Pitt, Chan, and Kohn}{Pitt
  et~al.}{2006}]{pitt2006efficient}
Pitt, M., D.~Chan, and R.~Kohn (2006).
\newblock Efficient {B}ayesian inference for {G}aussian copula regression
  models.
\newblock {\em Biometrika\/}~{\em 93\/}(3), 537--554.

\bibitem[\protect\citeauthoryear{Plummer, Best, Cowles, and Vines}{Plummer
  et~al.}{2006}]{codapkg}
Plummer, M., N.~Best, K.~Cowles, and K.~Vines (2006).
\newblock Coda: Convergence diagnosis and output analysis for mcmc.
\newblock {\em R News\/}~{\em 6\/}(1), 7--11.

\bibitem[\protect\citeauthoryear{Suchard, Lemey, Baele, Ayres, Drummond, and
  Rambaut}{Suchard et~al.}{2018}]{beast2018}
Suchard, M.~A., P.~Lemey, G.~Baele, D.~L. Ayres, A.~J. Drummond, and A.~Rambaut
  (2018).
\newblock Bayesian phylogenetic and phylodynamic data integration using {BEAST}
  1.10.
\newblock {\em Virus Evolution\/}~{\em 4\/}(1), vey016.

\bibitem[\protect\citeauthoryear{Tobin}{Tobin}{1958}]{tobin1958estimation}
Tobin, J. (1958).
\newblock Estimation of relationships for limited dependent variables.
\newblock {\em Econometrica: journal of the Econometric Society\/}, 24--36.

\bibitem[\protect\citeauthoryear{Tsionas and Michaelides}{Tsionas and
  Michaelides}{2016}]{tsionas2016spatial}
Tsionas, E.~G. and P.~G. Michaelides (2016).
\newblock A spatial stochastic frontier model with spillovers: Evidence for
  italian regions.
\newblock {\em Scottish Journal of Political Economy\/}~{\em 63\/}(3),
  243--257.

\bibitem[\protect\citeauthoryear{Wood}{Wood}{2017}]{mgcvcite}
Wood, S. (2017).
\newblock {\em Generalized Additive Models: An Introduction with R\/} (2 ed.).
\newblock Chapman and Hall/CRC.

\bibitem[\protect\citeauthoryear{Zareifard and Khaledi}{Zareifard and
  Khaledi}{2021}]{zareifard2021heterogeneous}
Zareifard, H. and M.~J. Khaledi (2021).
\newblock {A heterogeneous Bayesian regression model for skewed spatial data}.
\newblock {\em Spatial Statistics\/}~{\em 46}, 100545.

\bibitem[\protect\citeauthoryear{Zhang, Nishimura, Bastide, Ji, Payne, Goulder,
  Lemey, and Suchard}{Zhang et~al.}{2021}]{zhang2021large}
Zhang, Z., A.~Nishimura, P.~Bastide, X.~Ji, R.~P. Payne, P.~Goulder, P.~Lemey,
  and M.~A. Suchard (2021).
\newblock Large-scale inference of correlation among mixed-type biological
  traits with phylogenetic multivariate probit models.
\newblock {\em The Annals of Applied Statistics\/}~{\em 15\/}(1), 230--251.

\bibitem[\protect\citeauthoryear{Zhang, Nishimura, Trov{\~a}o, Cherry,
  Holbrook, Ji, Lemey, and Suchard}{Zhang et~al.}{2022}]{zhang2022hamiltonian}
Zhang, Z., A.~Nishimura, N.~S. Trov{\~a}o, J.~L. Cherry, A.~J. Holbrook, X.~Ji,
  P.~Lemey, and M.~A. Suchard (2022).
\newblock Accelerating bayesian inference of dependency between complex
  biological traits.
\newblock {\em arXiv preprint arXiv:2201.07291\/}.

\end{thebibliography}

\end{document}